\documentclass[aps,prb,floatfix,twocolumn,showpacs,preprintnumbers,amsmath,amssymb,superscriptaddress]{revtex4}

\usepackage{graphicx}  
\usepackage{dcolumn}   
\usepackage{bm}        
\usepackage{stmaryrd}  
\usepackage{multirow}  
\usepackage{color}     
\usepackage{amsmath,amssymb}
\usepackage{enumerate}

\newcommand{\der}[3]{\frac{{\rm d}^{#1} #2}{{\rm d} #3^{#1}}}     

\def   \bhs      {{\hat{\bm s}}}                         

\def   \Ms       {M_{\rm s}}                             

\def   \tq       {\tilde{q}}
\def   \txi      {\tilde{\xi}}
\def   \tg       {\tilde{g}}

\def   \e        {{\rm e}}                               
\def   \lsf      {l_{\rm sf}}                            
\def   \rsz      {\rho_0^{*}}                            
\def   \rs       {\rho^{*}}                              
\def   \rud      {\rho_{\uparrow (\downarrow)}}          
\def   \Rs       {R^{*}}                                 
\def   \Rsz      {R_0^{*}}                               
\def   \Rud      {R_{\uparrow (\downarrow)}}             
\def   \Gud      {G_{\uparrow\downarrow}}

\def   \DR       {\Delta R}                              

\def   \Heff     {{\bm H}_{\rm eff}}                     
\def   \hext     {H_{\rm ext}}                           
\def   \hani     {H_{\rm ani}}                           
\def   \Hdemag   {{\bm H}_{\rm dem}}                     
\def   \Hth      {{\bm H}_{\rm th}}                      

\def   \ez       {{\hat{\bm e}_{z}}}                     

\def   \gyro     {\gamma_{\rm g}}                        

\def   \nm       {{\rm nm}}                              

\def   \FL       {{\rm F_L}}
\def   \NL       {{\rm N_L}}
\def   \FC       {{\rm F_C}}
\def   \NR       {{\rm N_R}}
\def   \FR       {{\rm F_R}}

\begin{document}

\preprint{APS/123-QED}

\title{Nonlinear magnetotransport in dual spin valves}

\author{P.~Bal\'a\v{z}}
\affiliation{Department of Physics, Adam Mickiewicz University,
             Umultowska 85, 61-614~Pozna\'n, Poland}
\author{J.~Barna\'s}
\affiliation{Department of Physics, Adam Mickiewicz University,
             Umultowska 85, 61-614~Pozna\'n, Poland}
\affiliation{Institute of Molecular Physics, Polish Academy of Sciences
             Smoluchowskiego 17, 60-179 Pozna\'n, Poland}
\date{\today}

\begin{abstract}
Recent experiments on magnetoresistance in dual spin valves reveal
nonlinear features of electronic transport. We present a
phenomenological description of such nonlinear features
(current-dependent resistance and giant magnetoresistance) in
double spin valves. The model takes into account the dependence of
bulk/interface resistance and bulk/interface spin asymmetry
parameters for the central magnetic layer on charge current, and
consequently on spin accumulation. The model accounts for recent
nonlinear experimental observations.
\end{abstract}

\maketitle

\section{Introduction}

Spin accumulation (spin splitting of the
electrochemical potential) is a nonequilibrium phenomenon which is
associated with a spatially nonuniform spin asymmetry of two spin
channels for electronic transport~\cite{vanSon1987:PRL,JohnsonSilsbee1988:PRB,Barthelemy2002:JMMM}.
In the simplest case, it appears at the interface between ferromagnetic and non-magnetic
metals, when current has a nonzero component perpendicular to the
interface~\cite{Valet1993:PRB}. Spin accumulation also appears in
more complex systems, like single or double spin valves exhibiting
current-perpendicular-to-plane giant magnetoresistance (CPP-GMR)~\cite{Bass1999:JMMM,Katine2000:PRL}
effect, as well as in single or double tunnel junctions.
Current-induced spin accumulation is particularly pronounced in
spin-polarized transport through
nanoparticles~\cite{Barnas1998:PRL,Brataas1999:PRB} or quantum
dots and molecules~\cite{Rudzinski2001:PRB}.

In the case of spin valves based on layered magnetic structures,
spin accumulation and GMR are usually accounted for in terms of
the Valet-Fert description~\cite{Valet1993:PRB,Rychkov2009:PRL},
in which the spin accumulation is linear in current, while
resistance and magnetoresistance are independent of current
magnitude and current orientation. The description involves a
number of phenomenological parameters which usually are taken from
CPP-GMR experimental data. Originally, it was formulated for
collinear (parallel and antiparallel) magnetic configurations, but
later was extended to describe also current induced spin
torque~\cite{Barnas2005:PRB} and  CPP-GMR for arbitrary
noncollinear geometry~\cite{Gmitra2009:PRB}.

The Valet-Fert description was successfully applied not only to
single spin valves, but also to double (dual) spin
valves~\cite{Berger2003:JAP},  $\FL / \NL / \FC / \NR / \FR$,
where $\FC$ is a magnetically free layer separated from two
magnetically fixed outer layers ($\FL$ and $\FR$) by nonmagnetic
spacers ($\NL$ and $\NR$). An important feature of such structures
is an enhanced spin accumulation in the central layer ($\FC$) for
antiparallel  magnetizations of the outer magnetic layers (see
Fig.1). Spin accumulation may be then several times larger than in
the corresponding single spin valves. Accordingly, such a magnetic
configuration of dual spin valves (DSVs) diminishes the critical
current needed to switch magnetic moment of the central layer, and
also enhances the current-induced spin
dynamics~\cite{Berger2003:JAP,Balaz2009:PRB2}.
\begin{figure}[ht!]
 \centering
 \includegraphics[width=0.95\columnwidth]{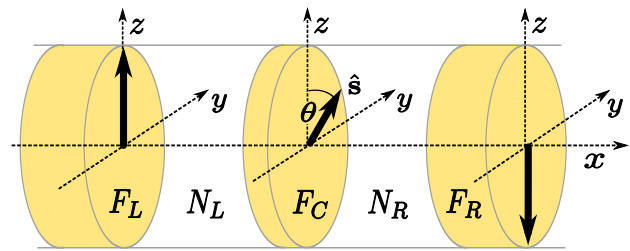}
 \caption{Scheme of a dual spin valve with antiparallel configuration of outer magnetic layers, $\FL$ and
 $\FR$. $\FC$ is the central magnetic layer .}
 \label{fig:model}
\end{figure}

Another interesting consequence of the enhanced spin accumulation
in the central layer of a dual spin valve is the possibility of
nonlinear transport effects. Recent experimental
results~\cite{Aziz2009:PRL} indicate that the enhanced spin
accumulation may cause unusual dependence of magnetoresistance on
dc current. It has been shown that when magnetizations of the
outer layers are antiparallel, resistance of a DSV for one current
orientation is lower when the $\FC$ layer is magnetized along the
$\FR$ one and higher when it is aligned along magnetization of the
$\FL$ layer, while for the opposite current orientation the
situation is reversed. Moreover, the difference in resistance of
both collinear configurations markedly depends on the applied
current. These observations strongly differ from the predictions
of the Valet-Fert model~\cite{Valet1993:PRB}, which gives
resistance (and magnetoresistance) independent of the current
density.

Such a nonlinear behavior may originate from several reasons. The
Valet-Fert description is based on the assumption of constant
(independent of spin accumulation and current) basic parameters of
the model, like bulk/interface resistance, bulk/interface spin
asymmetry, spin diffusion lengths, etc. This is justified when
spin accumulation is small and/or change in the density of states
on the energy scale comparable to spin accumulation is negligible
in the vicinity of the Fermi level. Density of states can be then
considered constant, i.e. independent of energy. Since the density
of states determines electron scattering rates, one may safely
assume that the transport parameters mentioned above are also
constant. However, when  the density of states at the Fermi level
varies remarkably with energy and spin accumulation is
sufficiently large, this assumption may not be valid, and the
parameters mentioned above may depend on spin
accumulation~\cite{Aziz2009:PRL}. This, in turn, may lead to
nonlinear effects, like the experimental ones described
above~\cite{Aziz2009:PRL}.

The spin accumulation, however, is rather small  -- of the order
of 0.1 meV for current density of  10$^8$ A/cm$^2$. Thus, to
account for the experimental observations one would need rather
large gradient of the density of states with respect to energy at
the Fermi level. More specifically, to account the experimental
observations, the change in density of states should be of the
order of 10\% on the energy scale of 1 meV. Although this is
physically possible, one cannot exclude other contributions to the
effect. Spin accumulation can directly change effective scattering
potential for electrons at the Fermi level. Moreover, spin
accumulation can also indirectly influence transport parameters,
for instance {\it via}  current-induced shift of the energy bands
due to charging of the layers or due to electron correlations,
which are neglected in the description of the spin accumulation.
Since the experimental results show that the nonlinear effects
appear only in the antiparallel configuration, where spin
accumulation in the central layer is large, we assume that the
indirect contributions are proportional to spin accumulation (at
least in the first order). Since, it is not clear which
contribution is dominant, we present a phenomenological approach,
which effectively includes all contributions to the observed
nonlinear transport. We assume that bulk and interfacial
resistances as well as spin asymmetries vary with spin
accumulation and show that such variation leads to effects
comparable to experimental observations~\cite{Aziz2009:PRL}.

Structure of this paper is as follows. In section II we describe
the model. Numerical results are presented in section III for bulk
and interfacial contributions. Section IV deals shortly with
magnetization dynamics in DSV. Finally, we conclude in section V.

\section{Model}

Electron scattering rate and its spin asymmetry become modified
when the spin-dependent Fermi levels are shifted (eg. due to spin
accumulation). All the effects leading to this modification can be
included in the description of charge and spin transport presented
in Refs.~\onlinecite{Barnas2005:PRB} and
\onlinecite{Gmitra2009:PRB}, which generalize the Valet-Fert model
to noncollinear magnetic configurations. We analyze the situation
when the effect originates from the bulk resistivity and bulk spin
asymmetry factor $\beta$ of the $\FC$ layer, which are assumed to
depend on spin accumulation, as well as from similar dependence of
the corresponding interface parameters. Let us begin with the bulk
parameters.

The spin-dependent bulk resistivity of a magnetic layer is
conveniently written in the form~\cite{Valet1993:PRB}
\begin{equation}
\label{Eq:bulkres}
  \rud = 2 \rs (1 \mp \beta)\, ,
\end{equation}
where $\rs$ is determined by the overall bulk resistivity $\rho_F$
as $\rs = \rho_F/(1-\beta^2)$. When the spin accumulation is
sufficiently large, one should take into account the corresponding
variation of $\rs$. In the lowest approximation (linear in the
spin accumulation) one can write
\begin{equation}
\label{Eq:rhovar}
  \rs = \rsz + q\, \langle g \rangle\, ,
\end{equation}
where $\rsz$ is the corresponding {\em equilibrium} (zero-current
limit) value, and $g(x)$ is spin accumulation in the central
layer, which varies with the distance from layer's interfaces. To
disregard this dependence, we average the spin accumulation over
the layer thickness $\langle g \rangle = (1 / d)\, \int_{\FC} g(x)
dx$. In Eq.(\ref{Eq:rhovar}) $q$ is a phenomenological parameter,
which depends on the relevant band structure. This parameter
effectively includes all effects leading to the modification of
transport parameters.

This equation can be rewritten as
\begin{equation}
\label{Eq:rhovar1}
  \rs = \rsz \, \left( 1 + \tq\, i\, \langle \tg \rangle \right)\, ,
\end{equation}
where $\tg$ is a dimensionless variable related to  spin
accumulation, $\tg = (\e^2 j_0 \rsz \lsf)^{-1}\, g$, with $j_0$
denoting the particle current density  and $\lsf$ being the
spin-flip length. We also introduced the dimensionless current
density $i = I / I_0$, with $I = \e j_0$ denoting the charge
current density and $I_0$ being a current density scale typical
for metallic spin valves ($I_0=10^8$A/cm$^2$). The parameter $\tq$
in Eq.~(\ref{Eq:rhovar1}), $\tq = (\e I_0 \lsf) q$, is a
dimensionless phenomenological parameter which is independent of
current.

The bulk spin asymmetry parameter $\beta$ becomes modified by spin
accumulation as well, and this modification can be written in a
form similar to that  in the case of $\rs$, i.e.
\begin{equation}
\label{Eq:betavar}
  \beta = \beta_0 + \xi \langle g \rangle\, ,
\end{equation}
where $\beta_0$ is the corresponding equilibrium value and $\xi$
effectively includes all the contributions. When introducing the
dimensionless spin accumulation defined above, one can rewrite
Eq.~(\ref{Eq:betavar}) as
\begin{equation}
\label{Eq:betavar1}
  \beta = \beta_0 + \txi\, i\, \langle \tg \rangle\, ,
\end{equation}
where $\txi = (e I_0 \rsz \lsf) \xi$.

Similar equations can be written for the interfacial resistance
$\Rs$ and interfacial asymmetry parameter $\gamma$, which define
spin-dependent interface resistance as
\begin{equation}
  \Rud = 2 \Rs (1 \mp \gamma)\, .
\end{equation}
Analogously, we can write the dependence of $\Rs$ and $\gamma$ on
spin accumulation in the form
\begin{subequations}
  \begin{align}
  \label{Eq:Rvar}
    \Rs    &= \Rsz + q'\, g(x_i)\, ,\\
  \label{Eq:gammavar}
    \gamma &= \gamma_0 + \xi'\, g(x_i)\, ,
  \end{align}
\label{Eqs:interface}
\end{subequations}
where $g(x_i)$ is spin accumulation at a given interface. The
constants $\Rsz$ and $\gamma_0$ are equilibrium interfacial
resistance and asymmetry parameter, respectively. Relations
(\ref{Eqs:interface}) lead to the following dependence of the
interfacial parameters on the current density:
\begin{subequations}
  \begin{align}
    \Rs &= \Rsz\, ( 1 + \tq'\, i\, \tg(x_i) )\, ,\\
    \gamma &= \gamma_0 + \xi'\, i\, \tg(x_i)\, ,
  \end{align}
\label{Eqs:interface1}
\end{subequations}
where $\tq' = (e I_0 \rsz \lsf)\, q'$  and $\txi' = (e I_0 \rsz
\lsf) \xi'$.

The parameters $q$, $\xi$, $q'$, and $\xi'$ introduced above
describe deviation from usual behavior of the resistance
(magnetoresistance) described by the Valet-Fert model. These
parameters will be considered as independent phenomenological
ones.

\section{Numerical results}

To calculate resistance and spin accumulation for arbitrary
noncollinear magnetic configuration, we apply the formalism
described in Refs.~\onlinecite{Barnas2005:PRB} and
\onlinecite{Gmitra2009:PRB}. This formalism, however, is modified
by assuming $\rs$, $\beta$, $\Rs$ and $\gamma$ to depend on
current density (spin accumulation). Therefore, for a particular
magnetic configuration and for certain values of  $i$, $\tq$,
$\txi$, $\tq'$, and $\txi'$, the spin accumulation has to be
calculated together with $\rs$, $\Rs$, $\beta$, and $\gamma$ in a
self-consistent way. In the first step, we assume equilibrium
values; $\rs = \rsz$ and $\beta = \beta_0$ ($\Rs = \Rsz$ and
$\gamma = \gamma_0$), and calculate the corresponding spin
accumulation $g_0(x)$ in the central magnetic layer. Then, we
calculate the zero approximation of the out-of-equilibrium
parameters according to Eqs. (\ref{Eq:rhovar1}),
(\ref{Eq:betavar1}), and/or (\ref{Eqs:interface1}). With these new
values for $\rs$ and $\beta$ ($\Rs$ and $\gamma$) we calculate the
out-of-equilibrium spin accumulation in the central layer and {\em
new} out-of-equilibrium values of $\rs$ and $\beta$ ($\Rs$ and
$\gamma$). The iteration process is continued until a stable point
is reached. Finally, for the obtained values of $\rs$, $\beta$,
$\Rs$, $\gamma$, and spin accumulation, we calculate the
resistance $R$ of the DSV at a given magnetic configuration (see
Ref.~\onlinecite{Gmitra2009:PRB}).

In all our calculations, magnetizations of the outermost layers
are assumed to be fixed and antiparallel (like in
experiment~\cite{Aziz2009:PRL}). Current is defined as positive
for electrons flowing from $\FR$ towards $\FL$. The equilibrium
parameters have been taken from the relevant literature (see
Appendix).

In this section we apply the above described model to two examples
of DSV structures. The first one is a symmetric DSV with $\FL =
\FR = {\rm Co}(20\nm)$, $\FC = {\rm Py}(8\nm)$, and with the
magnetic layers separated by $10\nm$ thick Cu spacers. The second
considered structure is an asymmetric exchange-biased DSV similar
to that used in experiment~\cite{Aziz2009:PRL}, namely
Cu-Co(6)/Cu(4)/Py(2)/Cu(2)/Co(6)/IrMn(10)-Cu, where the numbers in
 brackets correspond to layer thicknesses in nanometers.

\subsection{Bulk effects}

\begin{figure}[ht!]
 \centering
 \includegraphics[width=0.95\columnwidth]{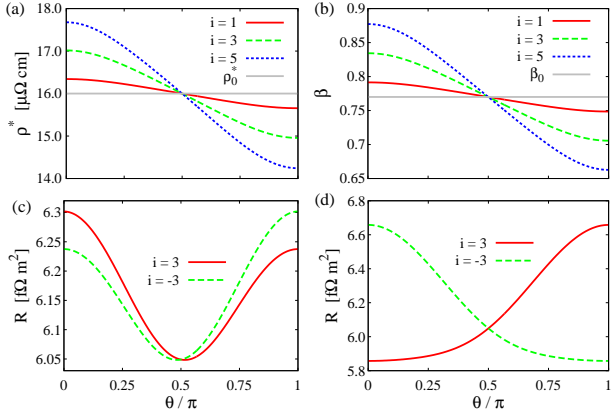}
 \caption{Symmetric dual spin valve
          Cu-Co(20)/Cu(10)/Py(8)/Cu(10)/Co(20)-Cu:
          (a) angular dependence of $\rs$ for $\tq = 0.1$ and $\txi = 0$;
          (b) angular dependence of $\beta$ for $\txi = 0.1$ and $\tq = 0$;
          (c) angular dependence of the resistance (per unit square) for $\tq = 0.1$ and $\txi = 0$;
          (d) angular dependence of the resistance for $\txi = 0.1$ and $\tq = 0$.
          The relative current density $i$ as indicated. }
 \label{fig:sym_bulk}
\end{figure}

In this subsection we consider pure bulk effects assuming $\tq' =
0$ and $\txi' = 0$. We start from  a symmetric DSV, and the
corresponding numerical results are shown in
Fig.~\ref{fig:sym_bulk}. First, we analyze the case with  $\tq =
0.1$ and $\txi = 0$. Figure~\ref{fig:sym_bulk}(a) shows how $\rs$
varies when magnetization of the central layer is rotated in the
layer plane. This rotation is described by the angle $\theta$
between magnetizations of the $\FL$ and $\FC$ layers. The higher
the current density, the more pronounced is the deviation of $\rs$
from its equilibrium value $\rsz$. The current-induced change in
$\rsz$ reaches maxima  when magnetic moment of the central layer
is collinear with those of the outer layers. These maxima are
different for the two opposite orientations of the  magnetic
moment of $\FC$ layer, as the corresponding spin accumulations are
different. For $\theta = \pi / 2$, however, one finds $\rs =
\rsz$. This is because spin accumulation vanishes then due to
opposite contributions of both interfaces (for symmetric DSVs).
Variation of $\rs$ in Fig.~\ref{fig:sym_bulk}(a) is shown only for
positive current, $i>0$. When current is negative, the change in
$\rs$ due to spin accumulation  changes sign (not shown), as also
can be concluded from Fig.~\ref{fig:sym_bulk}(c).

The current-induced angular dependence of $\rs$ makes the
resistance of the DSV dependent on the current density. As shown
in Fig.~\ref{fig:sym_bulk}(c), the angular dependence of the
resistance, becomes asymmetric, i.e. its magnitudes in the
opposite collinear states ($\theta = 0$ and $\pi$) are different.
Such an asymmetric angular dependence qualitatively differs from
that obtained from the Valet-Fert description, where the
resistance is symmetric. When magnetization of the central layer
switches (e.g. due to an applied magnetic field) from one
collinear state to the opposite one, one finds a drop (positive or
negative) in the resistance, defined as $\DR = R(\theta = \pi) -
R(\theta = 0)$. Moreover, when the current direction is reversed,
the corresponding drop in resistance also changes sign, as shown
in Fig~\ref{fig:sym_bulk}(c).

Let us consider now the situation where $\beta$ changes with the
spin accumulation (current), while  $\rho^*$ is constant, $\txi =
0.1$ and $\tq = 0$. General behavior of $\beta$ and of the
corresponding resistance with the angle $\theta$ is similar to
that discussed above (see Fig~\ref{fig:sym_bulk}(b,d)), although
the sign of the resistance drop for the current of a given
orientation is now opposite to that obtained in the case discussed
above, compare Figs~\ref{fig:sym_bulk}(c) and (d). Generally, the
sign of the drop in resistance may be controlled by the parameters
$\txi$ and $\tq$.

In real structures, however, both parameters, $\txi$ and $\tq$,
may be different from zero, and the observed behavior results from
interplay of the bulk and interface effects discussed above. To
show this, we consider now an asymmetric exchange-biased DSV
structure, Cu-Co(6)/Cu(4)/Py(2)/Cu(2)/Co(6)/IrMn(10)-Cu, similar
to that studied experimentally.

\begin{figure}[ht!]
 \centering
 \includegraphics[width=0.95\columnwidth]{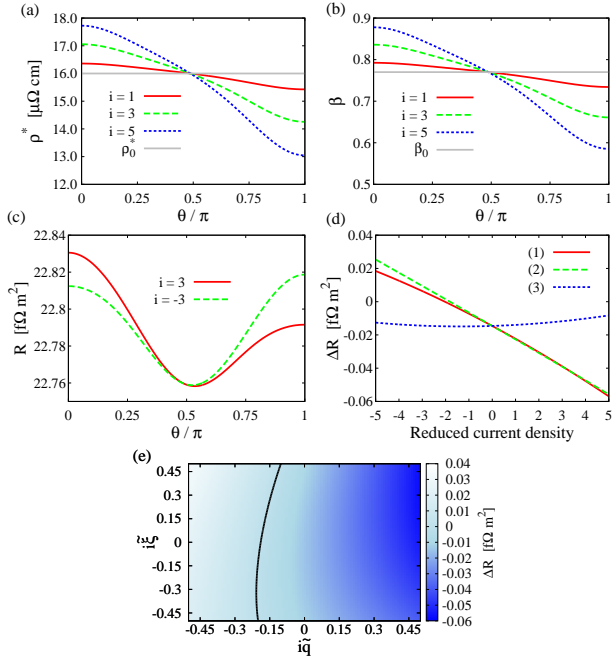}
 \caption{Asymmetric exchange-biased dual spin valve
          Cu-Co(6)/Cu(4)/Py(2)/Cu(2)/Co(6)/IrMn(10)-Cu:
          angular dependence of $\rs$ (a) and $\beta$ (b) for $\tilde{q} = 0.1$ and $\tilde{\xi} = 0.1$, and the
          angular dependence of the corresponding resistance (c);
          (d) dependence of the drops in resistance (per unit square) as a function of the reduced current density $i$
          for
              $\tilde{q} = 0.1$, $\tilde{\xi} = 0.1$ [line (1)],
              $\tilde{q} = 0.1$, $\tilde{\xi} = 10^{-3}$ [line (2)], and
              $\tilde{q} = 10^{-3}$, $\tilde{\xi} = 0.1$ [line (3)];
          (e) drop in the resistance as a function of
          $i\tq$ and $i\xi$ (with reduced current density $i$);
          the line covers the points where $\DR = 0$.
          }
 \label{fig:asym_bulk}
\end{figure}

Figures~\ref{fig:asym_bulk}(a) and (b) show the current-induced
angular dependence of $\rs$ and $\beta$ for  $\tq = \txi = 0.1$.
In comparison to the symmetric DSV structure, the difference in
the deviations of both parameters from their equilibrium values
for $\theta = \pi$ and $\theta = 0$ is now much more pronounced.
As before, the nonequilibrium values of the parameters cross the
corresponding equilibrium ones for nearly perpendicular
configuration, $\theta \approx \pi/2$. The resistance  shown in
Fig.~\ref{fig:asym_bulk}(c) reveals well defined drop between both
collinear configurations, and the  drop changes sign when the
current is reversed. However,  the resistance drops are now
different in their absolute magnitude due to the asymmetry of DSV.

Figure~\ref{fig:asym_bulk}(d) shows the resistance drops as a
function of the current density for three different sets of
parameters. For the parameters used in Fig. (3a-c), i.e. for $\tq
= \txi = 0.1$ [line (1)], the absolute value of the drop increases
rather linearly with increasing magnitude of current, although the
growth of $\DR$ is faster for positive than for negative current.
In the second case, $\tq = 0.1$ and $\txi = 10^{-3}$ [line (2)],
the dependence remains nearly the same, with only a small
deviation from the first case. Finally, we reduced the parameter
$\tq$, $\tq = 10^{-3}$, while $\txi = 0.1$ [line (3)]. Now, the
dependence strongly differs from the first two cases. $\DR$ only
slightly varies with current and remains rather small. Such a
behavior results from interplay of the bulk and interface
contributions. This interplay is presented also in
Fig.~\ref{fig:asym_bulk}(e), where the resistance drop is shown as
a function of $i\tq$ and $i\txi$. Additionally, the latter figure
shows that for any value of $\tq$ there is a certain value of
$\txi$ for which $\DR = 0$ (presented by the line).

\subsection{Interfacial effects}

\begin{figure}[ht!]
 \centering
 \includegraphics[width=0.95\columnwidth]{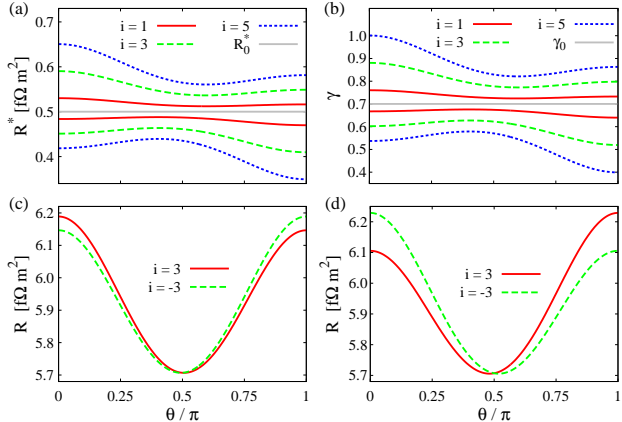}
 \caption{Symmetric dual spin valve
          Cu-Co(20)/Cu(10)/Py(8)/Cu(10)/Co(20)-Cu:
          (a) angular dependence of $\Rs$ on the left (curves below $\Rsz$)
              and right (curves above $\Rsz$) interfaces of the
              central layer
              for $\tq' = 0.1$ and $\txi' = 0$;
          (b) angular dependence of $\gamma$  on the left (curves below $\gamma_0$)
              and right (curves above $\gamma_0$) interfaces of
              the central layer
              for $\txi' = 0.1$ and $\tq' = 0$;
          (c) angular dependence of the resistance (per unit square) for $\tq' = 0.1$ and $\txi' = 0$;
          (d) angular dependence of the resistance (per unit square) for $\txi' = 0.1$ and $\tq' = 0$
          }
\label{fig:sym_interfaces}
\end{figure}

Now we consider the nonlinear effects due to current-dependent
interfacial parameters, as given by Eqs.~\ref{Eqs:interface1}. For
both  symmetric and asymmetric spin valves we  assume  that the
parameters $\tq'$ and $\txi'$ are equal for both interfaces of the
central layer. Consider first a symmetric DSV. The corresponding
results are summarized in Fig.~\ref{fig:sym_interfaces}. Variation
of $\Rs$, when the central magnetization rotates in the layer
plane, is shown in Fig.~\ref{fig:sym_interfaces}(a) for $\tq' =
0.1$ and $\txi' = 0$. The curves below the equilibrium value
$\Rsz$ correspond to $\Rs$ on the left interface, while these
above $\Rsz$ describe $\Rs$ on the right interface. When the
central magnetization is close to the collinear orientation
($\theta = 0$,$\pi$), $\Rs$ on the left and right interfaces are
significantly different, and this difference becomes partly
reduced when $\theta$ tends to  $\theta = \pi/2$ (for the systems
considered). Generally, the higher current density, the more
pronounced is the shift of $\Rs$ on both interfaces from their
equilibrium values. The corresponding angular dependence of the
DSV resistance is shown in Fig.~\ref{fig:sym_interfaces}(c) for
the current densities $I/ I_0 = \pm 3$. This angular dependence
results in small resistance drops of opposite signs for opposite
currents. As shall be shown below, the small value of $\DR$ is due
to a relatively large thickness of the central layer. Similar
conclusions can also be drown in the case when $\tq' = 0$ and only
$\gamma$ depends on spin accumulation, as shown in
Figs.~\ref{fig:sym_interfaces}(b) and (d).

\begin{figure}[ht!]
 \centering
 \includegraphics[width=0.95\columnwidth]{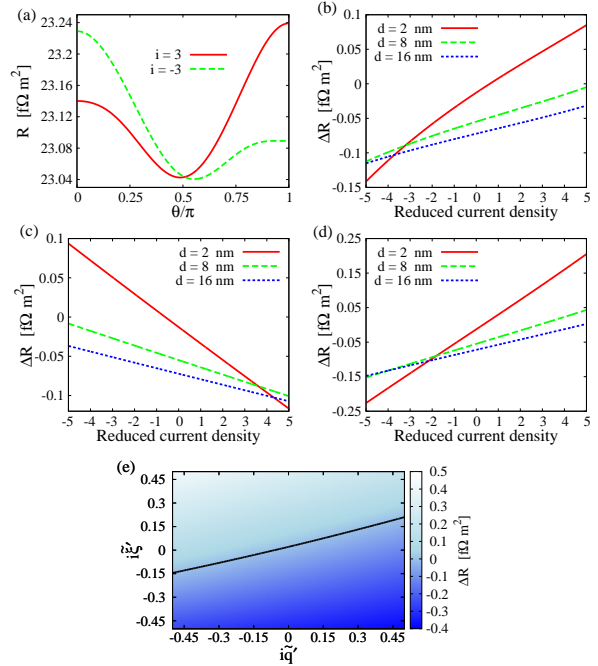}
 \caption{Asymmetric exchange-biased dual spin valve
          Cu-Co(6)/Cu(4)/Py($d$)/Cu(2)/Co(6)/IrMn(10)-Cu:
          (a) angular dependence of resistance (per unit square)
          calculated for central layer thickness $d = 2\,{\rm nm}$,
          $\tq' = 0.1$ and $\txi' = 0.1$;
          (b) dependence of the resistance drop (per unit square) on the reduced current density
          $i$ for $\tq' = 0.1$,  $\txi' = 0.1$, and  for different values of $d$;
          (c) resistance drop as a function of the reduced current density
          $i$ for $\tq' = 0.1$, $\txi' = 10^{-3}$, and indicated values of  $d$;
          (d) resistance drop {\it vs} current density
          $i$ for $\txi' = 0.1$, $\tq' = 10^{-3}$, and for indicated values of $d$;
          (e) resistance drop as a function of $i\tq'$ and $i\txi'$,
          calculated for $d = 2$. The
          line covers the points where $\DR = 0$.}
 \label{fig:asym_interfaces}
\end{figure}

For the asymmetric exchange-biased DSVs, we assume that both $\Rs$
and $\gamma$ depend on spin accumulation. As shown in
Fig.~\ref{fig:asym_interfaces}(a) for $\tq' = \txi' = 0.1$, there
is  a relatively large drop in resistance for the assumed
parameters. This resistance drop $\DR$ increases rather linearly
with the current density, as  shown in
Fig.~\ref{fig:asym_interfaces}(b). A small deviation from the
linear behavior can be observed only for larger values of negative
current. Calculations for different thicknesses of the central
layer, $d = 2$, $8$, and $16\,{\rm nm}$, show that the slope of
the curves representing the resistance drop as a function of the
current density decreases as the thickness increases, see
Fig.~\ref{fig:asym_interfaces}(b). In other words, the dependence
of resistance on current becomes less pronounced when the central
layer is thicker. We note, that such behavior was not observed in
the case of the bulk contribution. This feature arises because for
thicker magnetic layers the bulk resistivity dominates the pillar
resistance and suppresses the current-induced effects due to
interfaces. Additionally, the slope of the curves presenting the
resistance drop as a function of current density depends on the
parameters $\tq'$ and $\txi'$, and can change sign for appropriate
values of these parameters. This is shown in
Figs.~\ref{fig:asym_interfaces}(c) and (d), where one of the
parameters, either $\txi'$ (c) or $\tq'$ (d) has been reduced to
$10^{-3}$. Since $\tq'$ and $\txi'$ are of the same sign, their
effects are opposite and the corresponding contributions may
partly compensate each other. This is also shown in
Fig.~\ref{fig:asym_interfaces}(e), where the resistance drop $\DR$
is shown as a function of $i\tq'$ and $i\txi'$. From this figure
also follows that total compensation of the contributions to the
resistance drop occurs for the points corresponding to the line in
Fig.~\ref{fig:asym_interfaces}(e).

\section{Magnetization dynamics}

In the analysis presented above  magnetization of the central
layer was in the layer plane. However, when the magnetization
switches between the two collinear orientations (due to applied
magnetic field), it precesses and comes into out-of-plane
orientations as well. Such a precessional motion modifies spin
accumulation and also DSV's resistance. In this section we
describe variation of the resistance, when magnetization of the
central layer is switched by an external magnetic field back and
forth. To do this we make use of the single-domain approximation.
We also assume that the magnetic field is applied along the easy
axis of the central layer, similarly as in experiment (see
Fig.~\ref{fig:model}).

Time evolution of the spin moment of central layer is described by
the Landau-Lifshitz-Gilbert equation
\begin{equation}
  \der{}{\bhs}{t} = -|\gyro| \mu_0 \, \bhs \times \Heff + \alpha \, \bhs \times \der{}{\bhs}{t}\,
  .
\end{equation}
Here $\bhs$ is a unit vector along the spin moment of the central
layer, $\gyro$ is gyromagnetic ratio, $\mu_0$ vacuum permeability,
$\alpha$ is a dimensionless damping parameter, and $\Heff$ stands
for effective magnetic field,
\begin{equation}
   \Heff = -\hext \ez - \hani \left(\bhs \cdot \ez \right) \ez + \Hdemag + \Hth \, ,
\end{equation}
which includes external magnetic field ($\hext$) applied along
$\ez$-axis (see Fig.~\ref{fig:model}), anisotropy field ($\hani$),
and demagnetization field ($\Hdemag$) calculated for a layer of
thickness $d = 2\,{\rm nm}$ and elliptical shape with the major
and minor axes $130\,{\rm nm}$ and $60\,{\rm nm}$, respectively.
$\Hth$ is a stochastic gaussian field with dispersion $D = (\alpha
k_B T) / (\gyro \mu_0^2 \Ms V)$, which describes thermal
fluctuations at temperature $T$, where $k_B$ is the Boltzmann
constant, and $V$ is volume of the central magnetic layer.
Magnetic moments of the outer layers are assumed to be fixed due
to much larger coercive fields of these layers. Moreover, the
torque due to spin-transfer has not been included.
\begin{figure}[ht!]
 \centering
 \includegraphics[width=0.95\columnwidth]{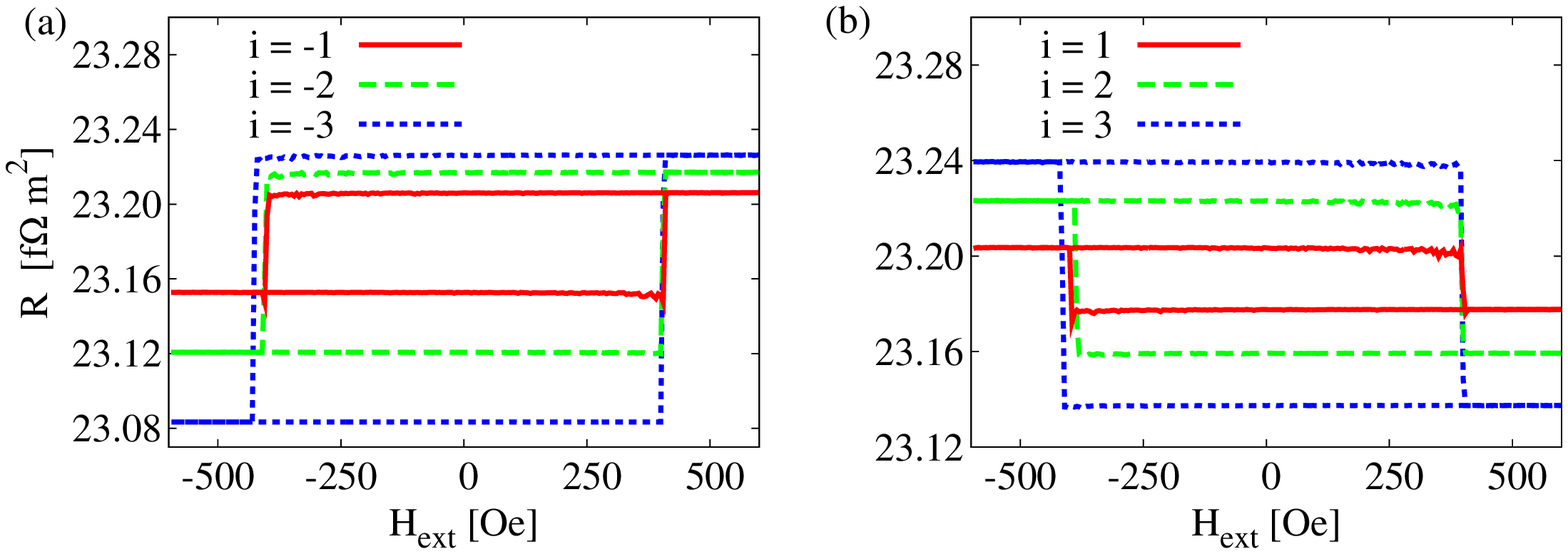}
 \caption{Asymmetric exchange-biased dual spin valve
          Cu-Co(6)/Cu(4)/Py(2)/Cu(2)/Co(6)/IrMn(10)-Cu:
          Minor hysteresis loops of resistance in external magnetic field
          calculated for $\tq'=0.1$, $\txi'=0.1$, and for different current densities $i$.
          Only interface contribution is considered here.}
 \label{fig:loops}
\end{figure}

Figure~\ref{fig:loops} shows quasistatic minor hysteresis loops of
the resistance in external magnetic field, calculated for
asymmetric exchange-biased DSV at $T = 70\,{\rm K}$. These figures
are in agreement with the results obtained in the preceding
section, and also in good agreement with experimental
observations~\cite{Aziz2009:PRL}. They also show that the drop in
resistance changes sign when the direction of current is reversed.
Moreover, one can observe small salient points in the hysteresis
loops, which appear during the reversal process -- especially at
low current densities. These points indicate on the minima in
resistance at noncollinear configurations and have been observed
experimentally as well.

The minor hysteresis loops appear also in the case when the
nonlinear effect is due to bulk parameters (not shown). Some
differences however appear, for instance in their dependence on
the layer thickness. This suggests, that the experimentally
observed effects are more likely due to interface contribution,
which is quite reasonable as the spin accumulation is maximal just
at the interfaces.

\section{Conclusions and discussion}

In summary, we have extended the description of spin accumulation
and magnetotransport in order to account for nonlinear
magnetotransport in metallic spin valves. We assumed the
dependence of bulk resistivities, interface resistances, and
bulk/interface asymmetry parameters on spin accumulation in the
central layer. The assumed phenomenological parameters effectively
include different contributions leading to modification of the
spin-dependent density of states at the Fermi levels. The obtained
numerical results reflect the trends observed experimentally. More
specifically, the dependence on spin accumulation of any of the
considered parameters leads to an asymmetric modification  of spin
valve resistance in comparison to its equilibrium (zero-current)
value. This modification  results in a drop in resistance when the
magnetic moment of the central layer switches between two
collinear configurations. Moreover, this drop depends on the
current density,  as has been also shown in
experiment~\cite{Aziz2009:PRL}. Within our phenomenological
description we can reproduce mainly linear dependence of the
current-induced resistance drops, with a small deviation from the
linearity for higher current densities. However, the description
fails to account strongly nonlinear variation of $\DR$, which was
observed in some DSV structures at high current
densities~\cite{Aziz2009:PRL}. To account for this behavior one
should take into account higher order terms in the expansion of
the relevant parameters. Additionally, when only interfacial
contribution is taken into account, the dependence of $\DR$ on
current becomes less pronounced with increasing thickness of the
central layer. Such a behaviour has been observed
experimentally~\cite{Aziz2009:PRL}, too, which indicates that the
interface contribution to the nonlinear effects is more important
than the bulk one.

The resistance drop measured experimentally at the current density
of $I = 10^{7}\, \mathrm{A cm^{-2}}$ is about $0.04\,
\mathrm{f\Omega m^{2}}$. To reach effects of similar magnitude
within the interfacial model, as shown in
Fig.~\ref{fig:asym_interfaces}, one needs $\txi' \sim 1$, i.e.
$\xi \simeq 1.13\, \mathrm{(meV)^{-1}}$ (when assuming the effect
is due to variation of interfacial asymmetry parameter only). If
direct contribution from spin accumulation would dominate, then
the corresponding change in the density of states would be of the
order of 10\%  on the energy scale of 1 meV. This slope may be
much smaller in the presence of other contributions.

\subsection*{Acknowledgment}

The work has been supported by the the EU through the Marie Curie
Training Network SPINSWITCH (MRTN-CT-2006-035327). Authors thank
Dr. Martin Gmitra for helpful discussions.

\appendix

\section{Parameters used in numerical calculations}

The parameters used in numerical calculations presented above have
been taken from \onlinecite{Bass1999:JMMM} and
\onlinecite{GmitraChapter}, and are summarized below:

\begin{table}[h!]
  \begin{center}
    \begin{tabular}{cccc}
    material & ~~$\rs$ [$\mu\Omega\, {\rm cm}$]~~ & ~~$\beta$~~ & ~~$\lsf$ [nm]~~\\
    \hline \hline
    Co & 5.1 & 0.51 & 60\\
    Py & 16 & 0.77 & 5.5\\
    Cu & 0.5 & 0 & 350\\
    IrMn & 150 & 0 & 1\\
    \hline
    \end{tabular}
  \end{center}
  \caption{Bulk parameters used in calculations.}
  \label{Tab:materials}
\end{table}

\begin{table}[h!]
  \begin{center}
    \begin{tabular}{ccccc}
    interface & ~~$\Rs$~~ & ~~$\gamma$~~ & ~~${\rm Re}\{\Gud\}$~~ & ~~${\rm Im}\{\Gud\}$~~\\
    \hline \hline
    Co/Cu & 0.5 & 0.77 & 0.542 & 0.016\\
    Py/Cu & 0.5 & 0.70 & 0.390 & 0.012\\
    Co/IrMn & 0.5 & 0.10 & -- & --\\
    IrMn/Cu & 0.5 & 0.70 & -- & --\\
    \hline
    \end{tabular}
  \end{center}
  \caption{Interfacial parameters used in calculations;
           $\Rs$ is given in ${\rm f}\Omega\, {\rm m}^2$, and mixing conductance,
           $\Gud$, in $({\rm f}\Omega\, {\rm m}^2)^{-1}$.
           Only parameters needed in calculations are given.}
\end{table}


\begin{thebibliography}{15}
\expandafter\ifx\csname natexlab\endcsname\relax\def\natexlab#1{#1}\fi
\expandafter\ifx\csname bibnamefont\endcsname\relax
  \def\bibnamefont#1{#1}\fi
\expandafter\ifx\csname bibfnamefont\endcsname\relax
  \def\bibfnamefont#1{#1}\fi
\expandafter\ifx\csname citenamefont\endcsname\relax
  \def\citenamefont#1{#1}\fi
\expandafter\ifx\csname url\endcsname\relax
  \def\url#1{\texttt{#1}}\fi
\expandafter\ifx\csname urlprefix\endcsname\relax\def\urlprefix{URL }\fi
\providecommand{\bibinfo}[2]{#2}
\providecommand{\eprint}[2][]{\url{#2}}

\bibitem[{\citenamefont{van Son et~al.}(1987)\citenamefont{van Son, van Kempe,
  and Wyder}}]{vanSon1987:PRL}
\bibinfo{author}{\bibfnamefont{P.~C.} \bibnamefont{van Son}},
  \bibinfo{author}{\bibfnamefont{H.}~\bibnamefont{van Kempe}},
  \bibnamefont{and} \bibinfo{author}{\bibfnamefont{P.}~\bibnamefont{Wyder}},
  \bibinfo{journal}{Phys. Rev. Lett.} \textbf{\bibinfo{volume}{58}},
  \bibinfo{pages}{2271} (\bibinfo{year}{1987}).

\bibitem[{\citenamefont{Johnson and Silsbee}(1988)}]{JohnsonSilsbee1988:PRB}
\bibinfo{author}{\bibfnamefont{M.}~\bibnamefont{Johnson}} \bibnamefont{and}
  \bibinfo{author}{\bibfnamefont{R.~H.} \bibnamefont{Silsbee}},
  \bibinfo{journal}{Phys. Rev. B} \textbf{\bibinfo{volume}{37}},
  \bibinfo{pages}{5326} (\bibinfo{year}{1988}).

\bibitem[{\citenamefont{Barthé\'el\'emy
  et~al.}(2002)\citenamefont{Barthé\'el\'emy, Fert, Contour, Bowen, Cros,
  Teresa, Hamzic, Faini, George, Grollier et~al.}}]{Barthelemy2002:JMMM}
\bibinfo{author}{\bibfnamefont{A.}~\bibnamefont{Barthé\'el\'emy}},
  \bibinfo{author}{\bibfnamefont{A.}~\bibnamefont{Fert}},
  \bibinfo{author}{\bibfnamefont{J.-P.} \bibnamefont{Contour}},
  \bibinfo{author}{\bibfnamefont{M.}~\bibnamefont{Bowen}},
  \bibinfo{author}{\bibfnamefont{V.}~\bibnamefont{Cros}},
  \bibinfo{author}{\bibfnamefont{J.~M.~D.} \bibnamefont{Teresa}},
  \bibinfo{author}{\bibfnamefont{A.}~\bibnamefont{Hamzic}},
  \bibinfo{author}{\bibfnamefont{J.~C.} \bibnamefont{Faini}},
  \bibinfo{author}{\bibfnamefont{J.~M.} \bibnamefont{George}},
  \bibinfo{author}{\bibfnamefont{J.}~\bibnamefont{Grollier}},
  \bibnamefont{et~al.}, \bibinfo{journal}{J. Magn. Magn. Mater}
  \textbf{\bibinfo{volume}{242}}, \bibinfo{pages}{68} (\bibinfo{year}{2002}).

\bibitem[{\citenamefont{Valet and Fert}(1993)}]{Valet1993:PRB}
\bibinfo{author}{\bibfnamefont{T.}~\bibnamefont{Valet}} \bibnamefont{and}
  \bibinfo{author}{\bibfnamefont{A.}~\bibnamefont{Fert}},
  \bibinfo{journal}{Phys. Rev. B} \textbf{\bibinfo{volume}{48}},
  \bibinfo{pages}{7099} (\bibinfo{year}{1993}).

\bibitem[{\citenamefont{Bass and Pratt}(1999)}]{Bass1999:JMMM}
\bibinfo{author}{\bibfnamefont{J.}~\bibnamefont{Bass}} \bibnamefont{and}
  \bibinfo{author}{\bibfnamefont{W.~P.} \bibnamefont{Pratt}},
  \bibinfo{journal}{J. Magn. Magn. Mater.} \textbf{\bibinfo{volume}{200}},
  \bibinfo{pages}{274} (\bibinfo{year}{1999}).

\bibitem[{\citenamefont{Katine et~al.}(2000)\citenamefont{Katine, Albert,
  Buhrman, Myers, and Ralph}}]{Katine2000:PRL}
\bibinfo{author}{\bibfnamefont{J.~A.} \bibnamefont{Katine}},
  \bibinfo{author}{\bibfnamefont{F.~J.} \bibnamefont{Albert}},
  \bibinfo{author}{\bibfnamefont{R.~A.} \bibnamefont{Buhrman}},
  \bibinfo{author}{\bibfnamefont{E.~B.} \bibnamefont{Myers}}, \bibnamefont{and}
  \bibinfo{author}{\bibfnamefont{D.~C.} \bibnamefont{Ralph}},
  \bibinfo{journal}{Phys. Rev. Lett.} \textbf{\bibinfo{volume}{84}},
  \bibinfo{pages}{3149} (\bibinfo{year}{2000}).

\bibitem[{\citenamefont{Barna\'s and Fert}(1998)}]{Barnas1998:PRL}
\bibinfo{author}{\bibfnamefont{J.}~\bibnamefont{Barna\'s}} \bibnamefont{and}
  \bibinfo{author}{\bibfnamefont{A.}~\bibnamefont{Fert}},
  \bibinfo{journal}{Phys. Rev. Lett.} \textbf{\bibinfo{volume}{80}},
  \bibinfo{pages}{1058} (\bibinfo{year}{1998}).

\bibitem[{\citenamefont{Brataas et~al.}(1999)\citenamefont{Brataas, Nazarov,
  Inoue, and Bauer}}]{Brataas1999:PRB}
\bibinfo{author}{\bibfnamefont{A.}~\bibnamefont{Brataas}},
  \bibinfo{author}{\bibfnamefont{Y.~V.} \bibnamefont{Nazarov}},
  \bibinfo{author}{\bibfnamefont{J.}~\bibnamefont{Inoue}}, \bibnamefont{and}
  \bibinfo{author}{\bibfnamefont{G.~E.~W.} \bibnamefont{Bauer}},
  \bibinfo{journal}{Phys. Rev. B} \textbf{\bibinfo{volume}{59}},
  \bibinfo{pages}{93} (\bibinfo{year}{1999}).

\bibitem[{\citenamefont{Rudzinski and Barna\'s}(2001)}]{Rudzinski2001:PRB}
\bibinfo{author}{\bibfnamefont{W.}~\bibnamefont{Rudzinski}} \bibnamefont{and}
  \bibinfo{author}{\bibfnamefont{J.}~\bibnamefont{Barna\'s}},
  \bibinfo{journal}{Phys. Rev. B} \textbf{\bibinfo{volume}{64}},
  \bibinfo{pages}{085318} (\bibinfo{year}{2001}).

\bibitem[{\citenamefont{Rychkov et~al.}(2009)\citenamefont{Rychkov, Borlenghi,
  Jaffres, Fert, and Waintal}}]{Rychkov2009:PRL}
\bibinfo{author}{\bibfnamefont{V.~S.} \bibnamefont{Rychkov}},
  \bibinfo{author}{\bibfnamefont{S.}~\bibnamefont{Borlenghi}},
  \bibinfo{author}{\bibfnamefont{H.}~\bibnamefont{Jaffres}},
  \bibinfo{author}{\bibfnamefont{A.}~\bibnamefont{Fert}}, \bibnamefont{and}
  \bibinfo{author}{\bibfnamefont{X.}~\bibnamefont{Waintal}},
  \bibinfo{journal}{Phys. Rev. Lett.} \textbf{\bibinfo{volume}{103}},
  \bibinfo{pages}{066602} (\bibinfo{year}{2009}).

\bibitem[{\citenamefont{Barna\'s et~al.}(2005)\citenamefont{Barna\'s, Fert,
  Gmitra, Weymann, and Dugaev}}]{Barnas2005:PRB}
\bibinfo{author}{\bibfnamefont{J.}~\bibnamefont{Barna\'s}},
  \bibinfo{author}{\bibfnamefont{A.}~\bibnamefont{Fert}},
  \bibinfo{author}{\bibfnamefont{M.}~\bibnamefont{Gmitra}},
  \bibinfo{author}{\bibfnamefont{I.}~\bibnamefont{Weymann}}, \bibnamefont{and}
  \bibinfo{author}{\bibfnamefont{V.}~\bibnamefont{Dugaev}},
  \bibinfo{journal}{Phys. Rev. B} \textbf{\bibinfo{volume}{72}},
  \bibinfo{pages}{024426} (\bibinfo{year}{2005}).

\bibitem[{\citenamefont{Gmitra and Barna\'s}(2009)}]{Gmitra2009:PRB}
\bibinfo{author}{\bibfnamefont{M.}~\bibnamefont{Gmitra}} \bibnamefont{and}
  \bibinfo{author}{\bibfnamefont{J.}~\bibnamefont{Barna\'s}},
  \bibinfo{journal}{Phys. Rev. B} \textbf{\bibinfo{volume}{79}},
  \bibinfo{pages}{012403} (\bibinfo{year}{2009}).

\bibitem[{\citenamefont{Berger}(2003)}]{Berger2003:JAP}
\bibinfo{author}{\bibfnamefont{L.}~\bibnamefont{Berger}}, \bibinfo{journal}{J.
  Appl. Phys.} \textbf{\bibinfo{volume}{93}}, \bibinfo{pages}{7693}
  (\bibinfo{year}{2003}).

\bibitem[{\citenamefont{Bal\'a\v{z} et~al.}(2009)\citenamefont{Bal\'a\v{z},
  Gmitra, and Barna\'s}}]{Balaz2009:PRB2}
\bibinfo{author}{\bibfnamefont{P.}~\bibnamefont{Bal\'a\v{z}}},
  \bibinfo{author}{\bibfnamefont{M.}~\bibnamefont{Gmitra}}, \bibnamefont{and}
  \bibinfo{author}{\bibfnamefont{J.}~\bibnamefont{Barna\'s}},
  \bibinfo{journal}{Phys. Rev. B} \textbf{\bibinfo{volume}{80}},
  \bibinfo{pages}{174404} (\bibinfo{year}{2009}).

\bibitem[{\citenamefont{Aziz et~al.}(2009)\citenamefont{Aziz, Wessely, Ali,
  Edwards, Marrows, Hickey, and Blamire}}]{Aziz2009:PRL}
\bibinfo{author}{\bibfnamefont{A.}~\bibnamefont{Aziz}},
  \bibinfo{author}{\bibfnamefont{O.~P.} \bibnamefont{Wessely}},
  \bibinfo{author}{\bibfnamefont{M.}~\bibnamefont{Ali}},
  \bibinfo{author}{\bibfnamefont{D.~M.} \bibnamefont{Edwards}},
  \bibinfo{author}{\bibfnamefont{C.~H.} \bibnamefont{Marrows}},
  \bibinfo{author}{\bibfnamefont{B.~J.} \bibnamefont{Hickey}},
  \bibnamefont{and} \bibinfo{author}{\bibfnamefont{M.~G.}
  \bibnamefont{Blamire}}, \bibinfo{journal}{Phys. Rev. Lett.}
  \textbf{\bibinfo{volume}{103}}, \bibinfo{pages}{237203}
  (\bibinfo{year}{2009}).
\end{thebibliography}

\end{document}